\documentclass[aps,prd,twocolumn,showpacs,showkeys,preprintnumbers,amsmath,amssymb]{revtex4-1}
\usepackage{makeidx}
\usepackage{amsfonts}
\usepackage{graphicx}
\usepackage{amsmath}
\usepackage{amssymb}

\begin{document}

\title{Teleparallel formalism of galilean gravity}
\author{S.C. Ulhoa }
\email{sc.ulhoa@gmail.com} \affiliation{Faculdade Gama,
\\Universidade de Bras\'{i}lia, 72405-610, Gama, DF, Brazil.}
\author{F.C. Khanna}
\email{fkhanna@ualberta.ca on} \affiliation{Physics Department,
Theoretical Physics Institute, University of Alberta
Edmonton, T6G 2J1, AB, Canada\\
TRIUMF 4004, Westbrook mall, Vancouver, British Columbia V6T 2A3,
Canada}
\author{A.E. Santana}
\email{asantana@fis.unb.br} \affiliation{Instituto de F\'isica and
International Center for Condensed Matter Physics, Universidade de
Bras\'ilia,  70910-900, Brasilia, DF,
Brazil\\
Physics Department, Theoretical Physics Institute, University of
Alberta Edmonton, T6G 2J1,  AB, Canada }

\begin{abstract}
A pseudo-Riemannian manifold is introduced, with light-cone
coordinates in (4+1) dimensional space-time, to describe a Galilei
covariant gravity. The notion of 5-bein and torsion are developed
and a galilean version of teleparallelism is constructed in this
manifold. The formalism is applied to two spherically symmetric
configurations. The first one is an ansatz which is inferred by
following the Schwarzschild solution in general relativity. The
second one is a solution of galilean covariant equations. In
addition, this Galilei teleparallel approach provides a prescription
to couple the 5-bein field to the galilean covariant Dirac field.
\end{abstract}



\maketitle

\section{Introduction}
\noindent

The Galilei symmetry provides the foundation of non-relativistic
classical and non-relativistic quantum mechanics, and phenomena in
many-body physics, such as superconductivity and nuclear systems,
are restricted to this regime~\cite{Fetter}. Beyond these, galilean
symmetry finds application even in the ultra-relativistic realm. It
has been observed that a Poincar\'e-covariant quantum field theory
in an infinite-momentum frame, in (3+1) dimensions, is reduced to a
Galilei covariant theory, in (2+1)
dimensions~\cite{PhysRev.150.1313,PhysRev.165.1535}. This is a
limiting process to describe particles with high velocities, which
is equivalent to taking the theory in the light-cone frame, where
the kinematical symmetry is the central extension of Galilei
group~\cite{PhysRevD.1.2795,PhysRevD.1.2901,Beckers1975338,%
PhysRevD.52.5954,PhysRevD.57.4965,PhysRevD.58.025013,PhysRevD.62.105015}.
In order to generalize this result to a Galilei symmetry in (3+1)
dimensions, the starting point is a theory  in a (4+1) space-time,
${\mathcal{G}}_{4,1}$. Then the Galilei symmetry is written in a
manifestly covariant form, and this formalism has been used in
various application~\cite{Montigny,Brihaye,Hagen}. In particular,
such a tensor structure has been associated with a non-relativistic
anti-de Sitter/conformal field theory (AdS/CFT), in order to
describe strongly coupled fermions, as it is the case of cold
fermion atoms at unitarity~\cite{conf1,conf2,conf3}.

The metric tensor for the galilean space-time is introduced by~\cite%
{Khanna,Santos}
\begin{equation}
\eta _{\mu \nu }=\left(
\begin{array}{ccc}
\mathbf{1}_{d} & 0 & 0 \\
0 & 0 & -1 \\
0 & -1 & 0%
\end{array}%
\right) \,,  \label{1}
\end{equation}%
where $\mathbf{1}_{d}$ is a Euclidian metric in $d$-dimensions. For $d=3,$
this metric leaves the scalar product of vectors invariant in the (4+1)
space-time. For instance, consider the 5-momentum $%
p=(p^{1},p^{2},p^{3},p^{4},p^{5})=(\mathbf{p},p^{4},p^{5})$, where $\mathbf{p%
}$ is the Euclidian linear momentum, $p^{4}$ is defined by the mass and $%
p^{5}$ is the energy. Using $\eta _{\mu \nu }$ and taking $p^{4}=m$ and $%
p^{5}=E/m$, up to a constant $v_{r}$ that is a scale of velocity
characteristic of the system, the Galilei invariant dispersion
relations reads $p^{2}=\mathbf{p}^{2}-2mE$. For $d=2$, the metric
$\eta _{\mu \nu }$ is associated with the light-cone coordinates.

This geometric structure leads to the construction of a galilean
covariant theory of gravitation~\cite{Ulhoa:2009at,Cuzinatto:2009dw}
based on a geometric approach. It provides solutions, for instance,
for the results of post-newtonian approximation of general
relativity, with no expansion in powers of $1/c$ while retaining
exact symmetry. Experiments, that have been suggested to test
Einstein theory of gravity, can be understood equally well within
this approach. This provides a strong motivation to pursue the
development of galilean symmetry and the theory of gravity, that can
also be useful in (2+1)-dimensional infinite-momentum frame
theories.

The galilean gravity has a symmetric connection, such that the
curvature ensures the dynamics of the field. On the other hand, this
suggests a formalism based in terms of a five-dimensional
Weitzenb\"{o}ck  space, where the curvature is identically zero and
the presence of the field is due to a non-null torsion,
corresponding to a 5-dimensional teleparallel approach.

In the context of the (3+1)-dimensional Poincar\'e symmetry, there
are two descriptions of gravity, the geometric approach and the
teleparallel formalism, which are equivalent to one another. The
teleparallelism is based on a set of 4-bein (vierbein) field, which
is introduced by describing the 4-dimensional space-time. Such a
field is related to the metric tensor of the
geometric formalism, by orthonormal relations~\cite%
{Hehl2,hehl-1994,Maluf2,MalufS,Aldrovandi}. There are at least two
appealing aspects to develop the teleparallelism. First, the
energy-momentum tensor for the Einstein gravitational theory is
defined without ambiguity. Second, a consistent definition of a
coupling between the gravitation and the Dirac fields is introduced.
An objective here is to show that both of these aspects are
established within a Galilei teleparallel formalism.

A proper definition for the energy-momentum tensor of the galilean
gravitational field is important for a comprehensive understanding
of systems in the weak-field regime. However, in this case, the
galilean gravity, developed as a geometric approach, shares the same
problem as the Einstein theory of general relativity: there is still
no agreement regarding an acceptable symmetric energy-momentum
tensor. Then we develop a teleparallel version of the galilean
gravity, based on the definition of a 5-bein field. As a result, we
 show that a symmetric energy-momentum tensor arises naturally from
the field equations. In addition, we consider a coupling of the
galilean gravitational field with spin-1/2 particles. The equation
of motion for the spin-1/2 particles is a manifestly
Galilei-covariant version of the Pauli-Schr\"odinger equation. There
is a lack of study in the literature about such a coupling and due
to the experimental interest this analysis is addressed here in
detail. In these developments the notion of Galilei covariance is
crucial.

The paper is organized as follows. In Section~\ref{II}, starting
with the metric $\eta_{\mu\nu}$, we introduce the 5-bein field and
establish the tangent space. In Section~\ref{III}, we derive a
teleparallel version field equations of the galilean gravity,
showing the equivalence between the geometric and the teleparallel
formulations. In addition, the energy-momentum tensor of the
galilean gravity is derived. We apply, in Section~\ref{IV}, the
formalism to two cases, both spherically symmetric systems that
represent the same configuration. In Section~\ref{V}, the coupling
of the galilean covariant Dirac field with the galilean gravity is
studied. In the final section we present some concluding remarks.
Appendix A gives a derivation of galilean covariant Dirac equation.
Appendix B provides a justification for a metric formulation of
galilean gravity, in particular the light-cone description. We use
natural units with $G=c=\hbar=1$.

\section{The 5-bein Field}

\label{II} \noindent

In the absence of the gravitational field, we consider physics as
being described in flat space-time, ${\mathcal{G}}_{4,1}$ (see the
Appendices A and B), whose metric is given in Eq.~(\ref{1}). Let us
designate quantities invariant under galilean transformations by
Latin indices. Such quantities will be called galilean vectors and
tensors. Those ones invariant under coordinate transformations will
be marked by using Greek indices. The local reference frame in
space-time adapted to an observer is realized by a set of orthogonal
5-vectors called the 5-bein field. It is represented by

\begin{equation}
e^a=e^a\,_\mu dx^\mu=(e^{(1)},e^{(2)},e^{(3)},e^{(4)},e^{(5)})\,,
\label{2.01}
\end{equation}
where the first three components are chosen following the
orientation of cartesian axes of the local reference frame. The
fourth component defines the field velocity of the observer and the
fifth one fixes the energy of such an observer. Hereafter we shall
use the term 5-bein field for $e^a\,_\mu$.

Considering a coordinate transformation, in flat space-time, $x^a\rightarrow x^\mu$ such as $%
x^{(1)}=r\sin\theta\cos\phi$, $x^{(2)}=r\sin\theta\sin\phi$,
$x^{(3)}=r\cos\theta$, $x^{(4)}=\frac{1}{\sqrt{2}}(x^{4}+x^{5})$ and
$x^{(5)}=\frac{1}{\sqrt{2}}(x^{4}-x^{5})$, the 5-bein field is
written as
\begin{small}
\begin{equation}
e^a\,_\mu = \frac{\partial x^a}{\partial x^\mu}= \left(
\begin{array}{ccccc}
\sin\theta\cos\phi & r\cos\theta\cos\phi & -r\sin\theta\sin\phi & 0 & 0 \\
\sin\theta\sin\phi & r\cos\theta\sin\phi & r\sin\theta\cos\phi & 0 & 0 \\
\cos\theta & -r\sin\theta & 0 & 0 & 0 \\
0 & 0 & 0 & \frac{\sqrt{2}}{2} & \frac{\sqrt{2}}{2} \\
0 & 0 & 0 & \frac{\sqrt{2}}{2} & -\frac{\sqrt{2}}{2} \\
&  &  &  &
\end{array}
\right).  \label{2.0}
\end{equation}
\end{small}
This coordinate transformation leads to a change in the galilean
metric from
$\eta_{ab}$ to the diagonal form $g_{\mu\nu}=diag(1,r^2,r^2\sin^2%
\theta,-1,1) $. Although the affine connection is different from
zero, the space remains flat since the curvature tensor is null. The
form given in
Eq.~(\ref{2.0}) will be used to construct the 5-bein field in section \ref%
{IV}.

A coordinate system $x^{\mu}$, in which gravitational forces are
present locally, can be realized as an inertial frame $x^a$ with
acceleration simulating gravitational interaction. At each
space-time point there is a flat tangent space. Greek indices run
from 1 to 5 and label world tensors on curved space, Latin indices
also run from 1 to 5 and denote galilean tensors on the tangent
space. The 5-bein field obeys the relations
\begin{eqnarray}
e^a\,_\mu e^b\,_\nu\eta_{ab}&=&g_{\mu\nu}\,,  \notag \\
e_a\,_\mu e_b\,_\nu g^{\mu\nu}&=&\eta_{ab}\,.  \label{2.1}
\end{eqnarray}
Therefore the galilean metric raises and lowers the Latin indices.
In a geometric description, gravity manifests itself by means of the
curvature tensor constructed from the affine connection while in
teleparallel version
it is realized through the torsion tensor constructed from the Weitzenb\"{o}%
ck like connection since the curvature tensor is zero in this case.

\section{Teleparallel Galilean Gravity}

\label{III} \noindent

Two vectors are said to be parallel if their projections on tangent space by
the action of the 5-bein field are equal. Thus, considering two vectors, $%
V^{a}(x)=e^{a}\,_{\mu }V^{\mu }(x)$ and $V^{a}(x+dx)=e^{a}\,_{\mu
}V^{\mu }(x)+(e^{a}\,_{\lambda }\partial _{\mu }V^{\lambda
}+V^{\lambda }\partial _{\mu }e^{a}\,_{\lambda })dx^{\mu }$,
separated from each other by an infinitesimal displacement, the
teleparallelism or distant parallelism is obtained if
$$
\nabla _{\nu }V^{\mu }=\partial _{\nu }V^{\mu }+(e^{a\mu }\partial _{\nu
}e_{a\lambda })V^{\lambda }=0\,,
$$%
where $e^{a\mu }\partial _{\nu }e_{a\lambda }=\Gamma _{\nu \lambda
}^{\mu }$ is called the Weitzenb\"{o}ck connection. The 5-bein field
satisfies the same condition.

It is well known that in the Riemannian geometry, the connection
(Christoffel symbols) is totally symmetric in its last two indices
which implies that the torsion tensor is zero. However,
Weitzenb\"{o}ck connection has a torsion tensor given by
\begin{equation}
T^{a}\,_{\mu\nu}(e)=\partial_\mu e^{a}\,_{\nu}-\partial_\nu
e^{a}\,_{\mu}\,, \label{3.0}
\end{equation}
and it is non-zero. On the other hand, the curvature tensor using
the Weitzenb\"{o}ck connection is identically zero. It is not
difficult to show that both connections are related by
\begin{equation}
\Gamma_{\mu ab}=^\circ\Gamma_{\mu ab}+K_{\mu ab}\,,  \label{3.01}
\end{equation}
where $K_{\mu ab}=\frac{1}{2}e_{a}\,^{\lambda}e_{b}\,^{\nu}
(T_{\lambda\mu\nu}+T_{\nu\lambda\mu}+T_{\mu\lambda\nu})$ is the contorsion
tensor and $^\circ\Gamma_{\mu ab}$ is the affine connection.

In order to establish the teleparallel version (distant parallelism) of
galilean gravity, it is worthwhile to show the relationship between the two
descriptions (geometric and teleparallel). Let us first write the curvature
scalar, $R(^{\circ }\Gamma )$, constructed from Eq. (\ref{3.01}). It is
possible to show that such a quantity is related to the curvature scalar
obtained from the affine connection in the following way
\begin{small}
\begin{equation}
eR(\Gamma )=eR(^{\circ }\Gamma )+e\left( {\frac{1}{4}}T^{abc}T_{abc}+{\frac{1%
}{2}}T^{abc}T_{bac}-T^{a}T_{a}\right) -2\partial _{\mu }(eT^{\mu })\,,
\label{3}
\end{equation}
\end{small}%
where $e$ is the determinant of the 5-bein field $e^{a}\,_{\mu }$ and $%
T^{a}=T^{b}\,_{b}\,^{a}$. Since the curvature tensor and all
contractions in teleparallelism are identically zero, it follows
that
\begin{small}
\begin{equation}
eR(^{\circ }\Gamma )\equiv -e\left( {\frac{1}{4}}T^{abc}T_{abc}+{\frac{1}{2}}%
T^{abc}T_{bac}-T^{a}T_{a}\right) +2\partial _{\mu }(eT^{\mu })\,.
\label{3.1}
\end{equation}
\end{small}%
Both sides of Eq. (\ref{3.1}) are invariant under galilean
transformations. Dropping the divergence term, the Lagrangian
density is given by
\begin{eqnarray}
\mathcal{L}(e_{a\mu})&=& -k\,e\,\left({\frac{1}{4}}T^{abc}T_{abc}+ {\frac{1}{%
2}} T^{abc}T_{bac} -T^aT_a\right) -{\mathcal{L}}_M  \notag \\
&\equiv&-k\,e \Sigma^{abc}T_{abc} -{\mathcal{L}}_M\;,  \label{3.2}
\end{eqnarray}
where $k=1/(16 \pi)$, ${\mathcal{L}}_M$ stands for the Lagrangian density
for the matter fields and $\Sigma^{abc}$ is defined by
\begin{equation}
\Sigma^{abc}={\frac{1}{4}} (T^{abc}+T^{bac}-T^{cab}) +{\frac{1}{2}}(
\eta^{ac}T^b-\eta^{ab}T^c)\;.  \label{3.3}
\end{equation}

Performing a variational derivative of the Lagrangian density with
respect to the 5-bein field $e_{a\lambda }$, we get the
Euler-Lagrange equation
\begin{equation}
\partial_\nu(e\Sigma^{a\lambda\nu})={\frac{1}{{4k}}} e\, e^a\,_\mu(
t^{\lambda \mu} + T^{\lambda \mu})\;,  \label{3.4}
\end{equation}
where
\begin{equation}
t^{\lambda \mu }=k(4\Sigma ^{bc\lambda }T_{bc}\,^{\mu }-g^{\lambda \mu
}\Sigma ^{bcd}T_{bcd})\,,  \label{3.5}
\end{equation}%
and $e^{a}\,_{\mu }T^{\lambda \mu }=\frac{1}{e}\delta
{\mathcal{L}}_{M}/e_{a\lambda }$ is the energy-momentum tensor of
the matter field. It is possible to show that the field equation
given in Eq. (\ref{3.4}) is equivalent to results in
the description of the geometric approach of the Galilei gravity~\cite%
{Ulhoa:2009at}. In addition, the Lagrangian densities are equivalent. Thus
it is clear that the teleparallel version of the galilean gravity and its
geometric description based on Riemannian geometry are indeed equivalent.

Now let us analyze the meaning of $t^{\lambda \mu }$. In view of the
antisymmetry property $\Sigma ^{a\mu \nu }=-\Sigma ^{a\nu \mu }$, it follows
that
\begin{equation}
\partial_\lambda \left[e\, e^a\,_\mu( t^{\lambda \mu} + T^{\lambda \mu})%
\right]=0\,,  \label{3.6}
\end{equation}
which is the local balance equation. Therefore we identify
$t^{\lambda\mu}$ as the galilean gravitational energy-momentum
tensor.

The integration of $t^{\lambda \mu }+T^{\lambda \mu }$ over a hyper-surface
of $x^{4}=constant$ defines the energy-momentum vector due to the galilean
gravitational and matter fields. Since the integrand is independent of $%
x^{(5)}$ for systems studied in this paper, the energy-momentum
vector reduces to
\begin{equation}
P^{a}=\int_{V}d^{3}x\,e\,e^{a}\,_{\mu }(t^{4\mu }+T^{4\mu })\,,  \label{3.8}
\end{equation}%
where $V$ is a volume of the three dimensional space. Using Eq. (\ref{3.4})
we obtain
\begin{equation}
P^{a}=\int_{V}d^{3}x\partial _{j}\Pi ^{aj}=\oint_{S}dS_{j}\,\Pi ^{aj}\,,
\label{3.9}
\end{equation}%
where $\Pi ^{aj}=4ke\,\Sigma ^{a4j}$. It is interesting to note that the
above expression is invariant under coordinate transformation and it is
projected on the tangent space, which means that it is frame dependent. In
such a formalism, the gravitation is considered as a manifestation of
torsion.

\section{Energy-Momentum Tensor of a Spherically Symmetric Configuration}

\label{IV} \noindent

There are two metrics describing spherical symmetry of a galilean gravity
field. One of them, analyzed in the following subsection, is obtained when a
similar form of Schwarzschild solution of general relativity is assumed~\cite%
{Ulhoa:2009at}. Another one arises from a direct solution of the
galilean field equations~\cite{Cuzinatto:2009dw}. Both cases explain
the advance of perihelion of Mercury, within galilei invariance and
without invoking Lorentz symmetry. We analyze both cases using the
energy-momentum tensor and considering the galilean teleparallelism.

\subsection{Case I}

\label{IV.I} \noindent

Consider the metric introduced in \cite{Ulhoa:2009at}. The line
element is

\begin{equation}
ds^{2}=f^{-1}(r)dr^{2}+r^{2}d\theta ^{2}+r^{2}\sin ^{2}\theta d\phi
^{2}-2f(r)(dx^{4})(dx^{5})\,,  \label{4}
\end{equation}%
where as usual $f=1-2M/r$. It describes a spherically symmetric
system and its form was set as an ansatz keeping in mind the
Schwarzschild solution in general relativity~\cite{Dinverno,Landau}.
In addition, it leads to the same result as for general relativity
for the case of the advance of perihelion of Mercury
\cite{Ulhoa:2009at}. Performing a coordinate transformation
\begin{equation}
x^{\prime 4}=\frac{1}{\sqrt{2}}(x^{4}+x^{5})\,,\ \ x^{\prime 5}=\frac{1}{%
\sqrt{2}}(x^{4}-x^{5})\,,  \notag \nonumber \label{4.1}
\end{equation}%
leaving the others coordinates unaltered, the line element assumes the form
\begin{equation}
ds^2=f^{-1}(r)dr^2+r^2d\theta^2+r^2\sin^2\theta d\phi^2-f(r){(dx^{\prime 4})}%
^2+f(r){(dx^{\prime 5})}^2\,.  \label{4.2}
\end{equation}

We choose a 5-bein field adapted to a spatial stationary frame which
means that $e_{(4)}\,^{\mu }=(0,0,0,c_{1},c_{2})$, where $c_{1}$ and
$c_{2}$ are constants. The interpretation of $e_{(4)}\,^{\mu }$ as
the field velocity is supported by the fact that such a component
lies along a tangent of the world line for an observer. In addition,
at spatial infinity, the axes of
the coordinate frame should be at rest, which is possible with the choices $%
e_{(1)}\,^{\mu }=(1,0,0,0,0)$, $e_{(2)}\,^{\mu }=(0,1,0,0,0)$ and $%
e_{(3)}\,^{\mu }=(0,0,1,0,0)$ in an asymptotic limit, using
cartesian coordinates. Therefore the general form of the 5-bein
field is
\begin{small}
\begin{equation}
e^{a}\,_{\mu }=\left(
\begin{array}{ccccc}
A\sin \theta \cos \phi & r\cos \theta \cos \phi & -r\sin \theta \sin \phi & 0
& 0 \\
A\sin \theta \sin \phi & r\cos \theta \sin \phi & r\sin \theta \cos \phi & 0
& 0 \\
A\cos \theta & -r\sin \theta & 0 & 0 & 0 \\
0 & 0 & 0 & \frac{\sqrt{2}B}{2} & \frac{\sqrt{2}C}{2} \\
0 & 0 & 0 & \frac{\sqrt{2}B}{2} & -\frac{\sqrt{2}C}{2}%
\end{array}%
\right) ,  \label{4.3}
\end{equation}
\end{small}%
where the functions $A$, $B$ and $C$ are defined by the condition $%
e^{a}\,_{\mu }e_{a}\,_{\nu }=g_{\mu \nu }$, yielding
\begin{eqnarray}
A^{2} &=&f^{-1}(r)\,,  \notag \\
B^{2} &=&f(r)\,,  \notag \\
C^{2} &=&f(r)\,.  \label{4.4}
\end{eqnarray}%
Then the determinant of the 5-bein field given in Eq. (\ref{4.3}) is $%
e=f^{1/2}r^{2}\sin \theta $.

In order to obtain the energy-momentum tensor, we perform the calculations
in Eq. (\ref{3.9}) on a spherical hyper-surface with infinite radius. We get
the following components of $\Sigma ^{\lambda \mu \nu }$, Eq. (\ref{3.3}),
\begin{eqnarray}
\Sigma ^{141} &=&\frac{1}{2}%
g^{11}g^{44}(g^{22}T_{224}+g^{33}T_{334}-g^{55}T_{545})\,,  \notag \\
\Sigma ^{241} &=&-\frac{1}{4}g^{22}g^{44}g^{11}(T_{214}+T_{412}+T_{124})\,,
\notag \\
\Sigma ^{341} &=&-\frac{1}{4}g^{33}g^{44}g^{11}(T_{314}+T_{413}+T_{134})\,,
\notag \\
\Sigma ^{441} &=&\frac{1}{2}%
g^{11}g^{44}(g^{22}T_{212}+g^{33}T_{313}+g^{55}T_{515})\,,  \notag \\
\Sigma ^{541} &=&-\frac{1}{4}g^{55}g^{44}g^{11}(T_{514}+T_{415}-T_{145})\,.
\label{4.5}
\end{eqnarray}%
The non-null components of the torsion tensor are
\begin{eqnarray}
T_{212}&=&r(1-A)\,,  \notag \\
T_{313}&=&T_{212}\sin^2\theta\,,  \notag \\
T_{515}&=&\frac{1}{2}\partial_1(C^2)\,.  \label{4.6}
\end{eqnarray}

Using Eq. (\ref{3.9}), Eq. (\ref{4.5}) and Eq. (\ref{4.6}), the
non-null components of the energy-momentum vector are
\begin{eqnarray}
P^{(4)} &=&\frac{\sqrt{2}M}{4}\,,  \notag \\
P^{(5)} &=&\frac{\sqrt{2}M}{4}\,.  \label{4.7}
\end{eqnarray}%
This result is expected since we are dealing with a frame which is
at rest. Therefore in such a frame $P^{(1)}$, $P^{(2)}$ and
$P^{(3)}$ should be null. The Casimir invariant is given by
$P^{2}=P^{a}P_{a}=-2P^{(4)}P^{(5)}$, i. e. it is $P^{2}=-M^{2}/4$.

\subsection{Case II}

\label{IV.II} \noindent

Let us now analyze the second case. The following line element
\begin{scriptsize}
\begin{equation}
ds^{2}=f^{-1}(r)dr^{2}+r^{2}d\theta ^{2}+r^{2}\sin ^{2}\theta d\varphi ^{2}+%
\frac{m}{r}(dx^{4})^{2}+\frac{m}{r}(dx^{5})^{2}-2f(r)(dx^{4})(dx^{5})
\label{4.8}
\end{equation}
\end{scriptsize}%
describes the invariant interval on the curved manifold with
spherical symmetry~\cite{Cuzinatto:2009dw}. It was obtained as an
exact solution of the field equations in galilean gravity and
applied to analyze the advance of the perihelion of Mercury and the
bending of light. Using the coordinate transformations $x^{\prime
4}=\frac{1}{\sqrt{2}}(x^{4}+x^{5})$ and $x^{\prime
5}=\frac{1}{\sqrt{2}}(x^{4}-x^{5})$, keeping the other variables the
same, the line element becomes
\begin{equation}
ds^2=f^{-1}(r)dr^2+r^2d\theta^2+r^2\sin^2\theta d\phi^2-f(r){(dx^{\prime 4})}%
^2+{(dx^{\prime 5})}^2\,,  \label{4.9}
\end{equation}
where $f=1-2M/r$.

Let us choose a 5-bein field adapted to a spatial stationary frame.
It has to obey the same conditions and has the form of Eq.
(\ref{4.3}), with the functions $A$, $B$ and $C$ chosen as
\begin{eqnarray}
A^{2} &=&f^{-1}(r)\,,  \notag \\
B^{2} &=&f(r)\,,  \notag \\
C &=&1\,.  \label{4.11}
\end{eqnarray}%
The determinant of the 5-bein field is $e=r^{2}\sin \theta $.

Again, we need the components $\Sigma^{\mu41}$ in order to calculate
the the energy-momentum vector, expressing them in terms of the
torsion tensor, we find
\begin{eqnarray}
\Sigma^{141}&=&\frac{1}{2}%
g^{11}g^{44}(g^{22}T_{224}+g^{33}T_{334}-g^{55}T_{545})\,,  \notag \\
\Sigma^{241}&=&-\frac{1}{4}g^{22}g^{44}g^{11}(T_{214}+T_{412}+T_{124})\,,
\notag \\
\Sigma^{341}&=&-\frac{1}{4}g^{33}g^{44}g^{11}(T_{314}+T_{413}+T_{134})\,,
\notag \\
\Sigma^{441}&=&\frac{1}{2}%
g^{11}g^{44}(g^{22}T_{212}+g^{33}T_{313}+g^{55}T_{515})\,,  \notag \\
\Sigma^{541}&=&-\frac{1}{4}g^{55}g^{44}g^{11}(T_{514}+T_{415}-T_{145})\,.
\label{4.12}
\end{eqnarray}
The only non-null components of the torsion tensor appearing in above
expressions are
\begin{eqnarray}
T_{212} &=&r(1-A)\,,  \notag \\
T_{313} &=&T_{212}\sin ^{2}\theta \,,  \label{4.13}
\end{eqnarray}%
which yield the following non-null components of $P^{a}$ when used in Eq. (%
\ref{3.9})
\begin{eqnarray}
P^{(4)} &=&\frac{\sqrt{2}M}{2}\,,  \notag \\
P^{(5)} &=&\frac{\sqrt{2}M}{2}\,.  \label{4.14}
\end{eqnarray}%
This result is also expected once the 5-bein field is chosen with
the frame
spatially at rest. The difference between the two cases is the factor $\frac{%
\sqrt{2}}{2}$ appearing in Eq. (\ref{4.14}), while in Eq. (\ref{4.7}) it is $%
\frac{\sqrt{2}}{4}$. Constructing $P^{2}$, from components
(\ref{4.14}), it is $-M^{2}$.

By means of the analysis of the Casimir invariant $P^{a}P_{a}=P^{2}$
for each case, we conclude that the second metric is more
appropriate to describe a spherically symmetric configuration in the
realm of galilean gravity.

\section{Coupled Galilean Gravity and Dirac Field}
\label{V}

Let us consider a galilean covariant Dirac field, ${\Psi }(x)$,
defined on the five-dimensional manifold
${\mathcal{G}}_{\mathrm{(4+1)}}$ with the galilean
metric~\cite{Kobayashi:2007jn,Montigny:2007bv}. Then the lagrangian
density for such a field is given by
\begin{equation}
{\mathcal{L}}_{M}=\overline{\Psi }(x)(i\gamma ^{a}\overset{\leftrightarrow }{%
\partial _{a}}-\mu )\Psi (x)\,\label{5}
\end{equation}%
where $\phi \overset{\leftrightarrow }{\partial _{\nu }}\psi \equiv \frac{1}{%
2}\left[ \phi \partial _{\nu }\psi -(\partial _{\nu }\phi )\psi
\right] $ and the adjoint field is defined as
$$
\overline{\Psi }(x)=\Psi ^{\dagger }(x)\ \gamma ^{0},\label{5.01}
$$%
where
$$
\gamma ^{(0)}=\frac{1}{\sqrt{2}}\left( \gamma ^{(4)}+\gamma
^{(5)}\right) \,.\label{5.1}
$$%
The field ${\Psi }(x)$ and its adjoint $\overline{\Psi }(x)$ obey
anticommutation relations. The quantity $\mu $ is the invariant
associated with the square momentum on
${\mathcal{G}}_{\mathrm{(4+1)}}$. Thus defining a 5-momentum as
$p_{a}=i\partial _{a}=(i\nabla ,i\partial _{4},i\partial
_{5})=(\mathbf{p},E,m),$ an invariant is $p^{2}=\eta
^{ab}p_{a}p_{b}=\mu ^{2} $. The matrices $\gamma ^{a}$  obey the
Clifford algebra
$$
\left\{ \gamma ^{a},\gamma ^{b}\right\} =2\eta ^{ab}.
$$%
In Appendix A the representations of  $\gamma ^{a}$ in 4-dimension
are given, for a  (3+1) dimensional space-time.

Let us apply the variational principle to the free Lagrangian (\ref{5}).
Then the Euler-Lagrange equations of motion for $\Psi (x)$ and its adjoint $%
\overline{\Psi }(x)$ are respectively
$$
\left( i\gamma ^{a}\partial _{a}-\mu \right) \Psi (x)=0\qquad \mathrm{and}%
\qquad \overline{\Psi }(x)(i\gamma ^{a}\overset{\leftarrow }{\partial }%
_{a}+\mu )=0,
$$%
where $\phi \overset{\leftarrow }{\partial }\psi =(\partial \phi )\psi $.

In order to couple the galilean gravity with the galilean-covariant
Dirac field, we make use of the Levi-Civita connection $^{\circ
}\omega _{\mu ab}$ which transforms as a vector under coordinate
transformations and as a galilean tensor of rank two. Such a
connection was introduced in the framework of teleparallelism
equivalent to the general relativity to couple gravitation in (3+1)
dimension with the Lorentz invariant Dirac field. It leads to the
correct field equations and to a vanishing skew-symmetric part of
the stress tensor of matter fields~\cite{1978GReGr...9..621C}.
Therefore we shall use the prescription $\partial _{a}\rightarrow
D_{a}\equiv e_{a}\,^{\mu }D_{\mu }$, where $D_{\mu }$ is given by
\begin{equation}
D_{\mu }=(\partial _{\mu }-\frac{\imath }{2}\,^0\omega_{\mu
}\,^{ab}\Sigma _{ab})\,, \label{5.2}
\end{equation}%
with $\Sigma _{ab}=\frac{\imath }{4}[\gamma ^{a},\gamma ^{b}]$ and

\begin{eqnarray}
{}^0\omega_{\mu ab}&=&-{1\over 2}e^c\,_\mu(
\Omega_{abc}-\Omega_{bac}-\Omega_{cab})\,, \\ \nonumber
\Omega_{abc}&=&e_{a\nu}(e_b\,^\mu\partial_\mu
e_c\,^\nu-e_c\,^\mu\partial_\mu e_b\,^\nu)\,,\label{5.301}
\end{eqnarray}
the Christoffel symbols ${}^0\Gamma^\lambda_{\mu\nu}$ and the
Levi-Civita connection are identically related by $\partial_\mu
e^a\,_\nu +
{}^0\omega_\mu\,^a\,_b\,e^b\,_\nu-{}^0\Gamma^\lambda_{\mu\nu}e^a\,_\lambda=0$,
or
$${}^0\Gamma^\lambda_{\mu\nu}=e^{a\lambda}\partial_\mu e_{a\nu}+
e^{a\lambda}\,(^0\omega_{\mu ab})e^b\,_\nu\,.$$ Taking into account
(\ref{3.01}) we get $^{\circ }\omega _{\mu ab}=-K_{\mu ab}$.
Therefore we can write
\begin{equation}
D_{\mu }=(\partial _{\mu }+\frac{\imath }{2}K_{\mu }\,^{ab}\Sigma
_{ab})\,. \label{5.4}
\end{equation}%

Thus applying this prescription, the modified lagrangian density for
the Dirac field is
\begin{equation}
{\mathcal{L}}_{M}=e\overline{\Psi }(x)(i\gamma ^{\nu }\overset{%
\leftrightarrow }{D_{\nu }}-\mu )\Psi (x)\,\label{5.3},
\end{equation}%
where $e$ is the determinant of 5-bein field and $\gamma ^{\mu
}=e_{a}\,^{\mu }\gamma ^{a}$. Applying the variational derivative
with respect to $\Psi (x)$ and to its adjoint respectively we get
$$
\left( i\gamma ^{\nu }D_{\nu }-\mu \right) \Psi (x)=0\qquad \mathrm{and}%
\qquad \overline{\Psi }(x)(i\gamma ^{\nu }\overset{\leftarrow }{D}_{\nu
}+\mu )=0.
$$%
These are the coupled galilean-covariant Dirac equation. Correct
field equations are obtained using the full Lagrangian density,
Eq.~(\ref{3.2}), using ${\mathcal{L}}_{M}$ given by Eq.~(\ref{5.3}).

\section{Conclusion}

Using light-cone coordinates in a (4+1)-dimension pseudo-Rimannian
manifold, $\mathcal{G}$, the Galilei symmetry, in (3+1)-dimensions,
is introduced in a covariant way. This manifold is used to define
5-bein fields in the tangent space. The manifestly Galilei-covariant
field equations for the teleparallel formalism of the galilean
gravity are derived. We show the equivalence of the teleparallelism
and the geometric approach. An energy-momentum tensor is obtained by
analyzing conserved quantities from the field equations.

We have developed a coupling of the galilean-covariant Dirac field
with the 5-bein field by introducing a covariant derivative
$\partial_a\rightarrow D_a$. We used a Levi-Civita connection
following Coote and Macfarlane~\cite{1978GReGr...9..621C} when they
treated the coupling of the Poincar\'e-covariant Dirac field with
gravity in the 5-bein formalism. They showed that the Levi-Civita
connection generates the correct field equations and is the only
possibility that leads to a symmetric energy-momentum tensor for the
Dirac field. The Galilei covariant form of the lagrangian density
for the Dirac field allows the use of the same prescription. It is
important to note that the introduction of the galilean metric
tensor for the space-time has been a central element to obtain these
results.

We apply the formalism to two cases of spherical symmetry and show,
by construction of the Casimir invariants, that the second line
element presented is more appropriate to describe such a
configuration. We address this problem from the point of view of the
teleparallel version of the galilean gravity where conserved and
frame dependent quantities are allowed to exist.

As a final observation, it is important to emphasize that the
galilean gravity reproduces the results of post-newtonian
approximation of general relativity, with no expansion in powers of
$1/c$, while maintaining the exact symmetry. And it is suitable for
an understanding of various experiments that have been considered to
establish the Einstein theory of gravity. A galilean-covariant
teleparallel formalism is relevant since the coupling of gravity
with spin 1/2 particles and the symmetric energy-momentum tensor are
introduced without ambiguity.
\newline

{\textbf{Acknowledgments}}: The authors thank NSERC from Canada,
CAPES and CNPq from Brazil for financial support.

\appendix
\section{Galilei-covariant  Dirac equation}\label{Ia}

In this appendix we describe the Lagrangian formalism for scalar and spinor
fields in a galilei covariant version. We use the metric given in Eq.~(\ref%
{1}) with a dispersion relation given as
$2mE={\mathbf{p}}^2+const.$.

The galilean covariance is introduced as an embedding of the
space-time point $(\mathbf{x},t)$ into a ($D+1$)-dimensional
space-time. We consider $D=4$, such that the coordinates are denoted as $%
x\equiv (x^{\mu})=(x^1, x^2, x^3, x^4, x^5) =(\mathbf{x}, t, s ) $
and the metric given as Eq.~(\ref{1}). This corresponds to a $4+1$
de Sitter space, ${\cal G}_{4,1}$, since the metric can be
diagonalized to diag$(1,1,1,-1,1)$. The scalar product
$(x|y)=\eta_{\mu\nu}x^\mu y^\nu=x^iy^i-x^5y^4-x^4y^5 \label{g2} $ is
invariant under linear transformations
$x^{\mu\prime}=\Lambda^{\mu\prime}_\nu x^{\nu}. $

A particular choice of $\Lambda$ is
\begin{equation}
\Lambda^\mu_\nu\equiv\left (
\begin{array}{ccccc}
R^1_1 & R^1_2 & R^1_3 & -V^1 & 0 \\
R^2_1 & R^2_2 & R^2_3 & -V^2 & 0 \\
R^3_1 & R^3_2 & R^3_3 & -V^3 & 0 \\
0 & 0 & 0 & 1 & 0 \\
-V_iR^i_1 & -V_iR^i_2 & -V_iR^i_3 & {\frac{1}{2}}\mathbf{V}^2 & 1%
\end{array}%
\right ).
\end{equation}
Then a 5-vector $x^\mu$ is transformed by
\begin{equation}
\begin{array}{lcl}
x^{\prime i} & = & R^i_jx^j-V^ix^4 \\
x^{\prime 4} & = & x^4 \\
x^{\prime 5} & = & x^5-V_i\left (R^i_jx^j\right )+{\frac{1}{2}}\mathbf{V}%
^2x^4%
\end{array}
\label{gvectortransf}
\end{equation}
where $i, j=1, 2, 3$ and $\mathbf{V}=(V_1, V_2, V_3)=(V^1, V^2, V^3)$ is the
relative velocity. These correspond to the Galilei homogeneous
transformations for the components $x^{\prime i}$, space, and $x^{\prime 4}$%
, time. The component $x^{\prime 5}$ is associated with the velocity
$V^i$, in the following sense. Considering the 5-vector in the
light-cone, we have
$$
dx^{\mu}dx_{\mu}=(d\mathbf{x})^2 -2dx^4dx_5=0,
$$
resulting in $dx_5=\mathbf{V}\cdot d\mathbf{x}/2$. Then the Galilei
invariant physics is derived by the embedding of the
(3+1)-dimensional space-time in the light-cone of the
(4+1)-dimensional de Sitter space-time.

An arbitrary Galilei-vector $A^\mu$, that transforms as $A^{\mu\prime}=%
\Lambda^{\mu\prime}_\nu A^\nu, $ under a Galilei boost transformation
(taking $\mathbf{R}=1$) reads
\begin{equation}
\begin{array}{lcl}
A^{i\prime} & = & A^i-V^iA^4 \\
A^{4\prime} & = & A^4 \\
A^{5\prime} & = & A^5-V^iA^i+{\frac{1}{2}}V^iV^iA^4%
\end{array}
\label{explicitgeneralgvectortransf}
\end{equation}
Basic vectors are then $x^\mu=({\mathbf{x}}, t, s)$ and its conjugate
momentum $p^\mu=(\mathbf{p}, p^4, p^5)=({\mathbf{p}}, m, E)$. The
transformation of $p$ leads to
$$
\begin{array}{l}
{\mathbf{p}}^\prime={\mathbf{R}}{\mathbf{p}}-m\mathbf{V} \\
m^\prime=m \\
E^\prime=E-({\mathbf{R}}{\mathbf{p}})\cdot\mathbf{V}+{\frac{1}{2}}m\mathbf{V}%
^2%
\end{array}%
$$
The corresponding five-gradient is $%
\partial_\mu=(\nabla, \partial_t, \partial_s)$ and using the galilean metric,
Eq.~(\ref{1}), we have: $x_\mu=({\mathbf{x}}, -s, -t),
\partial^\mu=(\nabla, -\partial_s, -\partial_t)$ and
$p_\mu=({\mathbf{p}}, -E, -m). $ For a
quantum theory, using the Dirac correspondence procedure, we have ${\hat p}%
_\mu=\left ({\hat{\mathbf{p}}}, -{\hat E}, -{\hat m}\right )\equiv
-i\hbar\partial_\mu=(-i\hbar\nabla, -i\hbar\partial_t, -i\hbar\partial_s)$.

Let us investigate isometries in $\mathcal{G}$. We show that the ten
dimensional Galilei algebra is a subalgebra of the fifteen dimensional set
of affine transformations in $\mathcal{G}$, i.e. $x^\mu\rightarrow
\Lambda^\mu_\nu x^\nu+a^\mu $ which leaves the scalar product, defined with
the metric $\eta_{\mu\nu}$, invariant, and such that $|\Lambda|=1$. The Lie
algebra of this group is generated by $\{M_{\mu\nu}, P_\mu\}$, where $%
\mu,\nu =1,\dots, 5$ and is analogous to the Poincar\'e algebra, but in a $%
4+1$ dimensional space. Its commutation relations are
$$
\begin{array}{rcl}
\left [M_{\mu\nu}, M_{\rho\sigma}\right ] & = & i\left
(\eta_{\mu\rho}M_{\nu\sigma}+\eta_{\nu\sigma}M_{\mu\rho}-
\eta_{\nu\rho}M_{\mu\sigma}-\eta_{\mu\sigma}M_{\nu\rho}\right ) \\
{\left [P_{\mu}, M_{\rho\sigma}\right ]} & = & -i\left
(\eta_{\mu\rho}P_\sigma-\eta_{\mu\sigma}P_\rho\right ) \\
{\left [P_\mu, P_\nu\right ]} & = & 0.%
\end{array}%
$$
The generators of the Galilei group are given by
$$
\begin{array}{l}
M_{ij}\rightarrow \epsilon_{ijk}J_k \\
M_{5i}=-M_{i5}\rightarrow K_i \\
P_4\rightarrow -H \\
P_i\rightarrow P_i \\
P_5\rightarrow -m\mathbf{1}%
\end{array}
\label{galileinaffinetransf}
$$
where we have added the central extension $m$ through the generator
$P_5$, which is another way to define mass as a relic of the fifth
dimension. Now we study the unitary scalar and spinor
representations of this algebra.

For a scalar representation, the starting point is the Casimir invariant,${%
\hat p}^2=k'^2$, that leads to the equation ${\hat p}^2\psi(x)=k'^2
$. Observe this is consistent with the mass-shell condition
\begin{equation}
 p^2={\mathbf p}^2-2mE=k'^2\,.\label{seg01.111}
\end{equation}
 The constant $k'$ may be
taken, without loss of generality, to be zero. Then we obtain a
galilean Klein-Gordon-like equation,
\begin{equation}
\partial^\mu\partial_\mu\psi(x)=\eta^{\mu\nu}\partial_\mu\partial_\nu \psi
(x)=0  \label{gkleingordonequation}
\end{equation}
which becomes,
\begin{equation}
\nabla^2\psi (x)-2\partial_5\partial_t\psi (x)=0.  \label{gkg2}
\end{equation}
Since ${\hat p}_5=-m$ is an invariant, we have $\partial_5 \psi (x)=-\frac{im%
}\hbar\psi (x). $ A solution is
$$
\psi (x)=\exp\left (-\frac{ims}\hbar\right )\psi (\mathbf{x}, t).
$$
Then we obtain the Schr\"odinger equation,
$$
\nabla^2\psi (\mathbf{x}, t)-2\left (-\frac{im}\hbar\right ) \partial_t\psi (%
\mathbf{x}, t)=0
$$
In this case we have the five-current which is a Galilei-vector $({\mathbf{\
J}}(x), J_0(x), {\mathcal{E}}(x))$ such that
\begin{equation}
\begin{array}{lcl}
{\mathbf{J}}(x) & = & -\frac i2\hbar [\psi^\dagger (x)\nabla\psi (x) -
(\nabla\psi^\dagger (x) )\psi (x) ] \\
{J_0(x)} & = & m\psi^\dagger (x)\psi (x) \\
{\mathcal{E}}(x) & = & \frac{\hbar^2}{2m}\nabla\psi^\dagger
(x)\cdot\nabla\psi (x) .%
\end{array}%
\end{equation}

A spin 1/2 representation is derived by setting $( \gamma^\mu\partial_\mu
+k)\Psi=0 $, that in the momentum space reads $( \gamma^\mu p_\mu
-ik)\Psi=0. $ The $\gamma$-matrices satisfy the Clifford algebra
\begin{equation}
\{\gamma^\mu,
\gamma^\nu\}=\gamma^\mu\gamma^\nu+\gamma^\nu\gamma^\mu=2\eta^{\mu\nu}
\end{equation}
In (4+1) dimension, a $4\times4$ representation for the $\gamma$
matrices is
\begin{equation}
\begin{array}{cc}
\gamma^1=\left (
\begin{array}{cccc}
0 & 1 & 0 & 0 \\
1 & 0 & 0 & 0 \\
0 & 0 & 0 & -1 \\
0 & 0 & -1 & 0%
\end{array}
\right), & \gamma^2=\left (
\begin{array}{cccc}
0 & -i & 0 & 0 \\
i & 0 & 0 & 0 \\
0 & 0 & 0 & i \\
0 & 0 & -i & 0%
\end{array}
\right) \\
\gamma^3=\left (
\begin{array}{cccc}
1 & 0 & 0 & 0 \\
0 & -1 & 0 & 0 \\
0 & 0 & -1 & 0 \\
0 & 0 & 0 & 1%
\end{array}
\right), & \gamma^4=\left (
\begin{array}{cccc}
0 & 0 & 0 & 0 \\
0 & 0 & 0 & 0 \\
-\sqrt{2} & 0 & 0 & 0 \\
0 & -\sqrt{2} & 0 & 0%
\end{array}
\right) \\
\gamma^5=\left (
\begin{array}{cccc}
0 & 0 & \sqrt{2} & 0 \\
0 & 0 & 0 & \sqrt{2} \\
0 & 0 & 0 & 0 \\
0 & 0 & 0 & 0%
\end{array}
\right). &
\end{array}
\label{diracmatrices}
\end{equation}

The adjoint spinor is defined as $\bar{\Psi} = \Psi^\dagger\zeta$ with
\begin{equation}
\zeta =\frac{-1}{\sqrt{2}}\left (\gamma^4+\gamma^5 \right )= \left (
\begin{array}{cccc}
0 & 0 & -1 & 0 \\
0 & 0 & 0 & -1 \\
1 & 0 & 0 & 0 \\
0 & 1 & 0 & 0%
\end{array}
\right ).  \label{zeta}
\end{equation}
The current is given by
\begin{equation}
j_{\mathrm{Dirac}}^\mu=\frac{-i}{\sqrt{2}}\bar{\Psi}\gamma^\mu\Psi .
\end{equation}
Using $( \gamma^\mu\partial_\mu -k )$, we get $(\eta^{\mu\nu}\partial_\mu%
\partial_\nu-k^2)\Psi=0 $ which (for $k=0$) is the Galilei
Klein-Gordon like equation.

The Lagrangian density for the galilean-Dirac equation is
$$
{\mathcal{L}}_{\mathrm{Dirac}}=\bar{\Psi} ( \gamma^\mu\partial_\mu + k)\Psi .
$$
Writing the spinor field as
\begin{equation}
\Psi=\left (%
\begin{array}{c}
\varphi \\
\chi%
\end{array}
\right )
\end{equation}
we find
\begin{equation}
\begin{array}{l}
\left(\sigma\cdot\mathbf{p}-ik \right) \varphi -\left(-im\omega\sigma\cdot%
\mathbf{r} +\sqrt{2}m\right) \chi =0 \\
-\left (\sigma\cdot\mathbf{p}+ik\right )\chi + \left(\sqrt{2}%
E+im\omega\sigma\cdot\mathbf{r} \right)\varphi =0.%
\end{array}
\label{e24}
\end{equation}
Subtracting the second equation from the first, we obtain
\begin{equation}
\left (\sigma\cdot\mathbf{p}_--ik\right )\varphi +\left (\sigma\cdot\mathbf{p%
}_++ik\right )\chi -\sqrt{2}E\varphi -\sqrt{2}m\chi =0.  \label{e25}
\end{equation}
If we express $\chi$ in terms of $\varphi$ as
\begin{equation}
\chi =\frac{\sqrt{2}}{2m}\left (\sigma\cdot\mathbf{p}-ik\right )\varphi
\label{e26}
\end{equation}
we find that, with the non-minimal substitution
\begin{equation}
\mathbf{p}\rightarrow \mathbf{p}-im\omega\zeta\mathbf{r},
\end{equation}
the relation between $\chi$ and $\varphi$ is
\begin{equation}
\chi =\frac{\sqrt{2}}{2m}\left(\sigma\cdot\mathbf{p}_--ik\right )\varphi
\label{e27}
\end{equation}
where we use the notation $\mathbf{p}_{\pm}\equiv\mathbf{p}\pm im\omega%
\mathbf{r}$. Then $\varphi$ satisfies the equation
\begin{equation}
\frac{\sqrt{2}}{2m}\left (\sigma\cdot\mathbf{p}_++ik\right )\left
(\sigma\cdot\mathbf{p}_--ik\right ) \varphi -\sqrt{2}E\varphi =0  \label{e29}
\end{equation}
from which we obtain
\begin{small}
\begin{equation}
E\varphi =\left( \frac{\mathbf{p}^2}{2m}+\frac{m\omega^{2}\mathbf{r}^{2}}{2}%
- \frac{3\omega}{2}-\frac{ik}{2m}\left(\sigma\cdot\mathbf{p}_+- \sigma\cdot%
\mathbf{p}_-\right )+\frac{k^{2}}{2m}\right)\varphi  \label{e30}
\end{equation}
\end{small}
or
$$
E\varphi =\left( \frac{\mathbf{p}^2}{2m}+\frac{m\omega^{2}\mathbf{r}r^{2}}{2}%
- \frac{3\omega}{2}+k\omega\sigma\cdot\mathbf{r}+\frac{k^{2}}{2m}\right)
\varphi\
$$
By redefining $E\rightarrow E+\frac{k^{2}}{2m},$ we have
\begin{equation}
E\varphi =\left( \frac{\mathbf{p}^2}{2m}+\frac{m\omega^{2}\mathbf{r}^{2}}{2}%
- \frac{3\omega}{2}+k\omega\sigma\cdot\mathbf{r}\right) \varphi .
\label{e32}
\end{equation}
This equation therefore describes a non-relativistic harmonic oscillator.

For a Galilei theory on the light-cone of the (3+1) Minkowski
space-time, the vectors are written as $x\equiv (x^{\mu})=(x^1, x^2,
x^-, x^+) =(\mathbf{x}, x^-, x^+ ) $ and the metric is given by
Eq.~(\ref{1}) with $d=2$. The covariant equations we have derived
for the scalar and spinor representations are recovered by the
change of notations, $x^4\rightarrow x^- $ and $x^5\rightarrow x^+$.
For the Dirac equation, the Dirac algebra is generated by the four
$\gamma$-matrices,
\begin{equation}
\begin{array}{cc}
\gamma^1=\left (
\begin{array}{cccc}
0 & 1 & 0 & 0 \\
1 & 0 & 0 & 0 \\
0 & 0 & 0 & -1 \\
0 & 0 & -1 & 0%
\end{array}
\right), & \gamma^2=\left (
\begin{array}{cccc}
0 & -i & 0 & 0 \\
i & 0 & 0 & 0 \\
0 & 0 & 0 & i \\
0 & 0 & -i & 0%
\end{array}
\right) \\
\gamma^-=\left (
\begin{array}{cccc}
0 & 0 & \sqrt{2} & 0 \\
0 & 0 & 0 & \sqrt{2} \\
0 & 0 & 0 & 0 \\
0 & 0 & 0 & 0%
\end{array}
\right), & \gamma^+=\left (
\begin{array}{cccc}
0 & 0 & 0 & 0 \\
0 & 0 & 0 & 0 \\
-\sqrt{2} & 0 & 0 & 0 \\
0 & -\sqrt{2} & 0 & 0%
\end{array}
\right) .%
\end{array}%
\end{equation}

\section{Metric Formulation of Galilean Gravity}

The purpose of this appendix is to recall the ideas developed in
Ref. \cite{Ulhoa:2009at} about the geometric approach to galilean
gravity. Thus we start with the Galilei transformations written in a
covariant form as described in the last appendix and construct a
galilean theory of gravity by introducing a Riemannian manifold
where locally we have the (4+1)-space-time, ${\cal G}_{4,1}$. We
carry out the reduction from the (4+1) dimensional space-time to the
(3+1) dimensional manifold in order to provide physical
interpretation of field equations.

Taking a non-flat manifold where locally the metric is
$\eta_{\mu\nu}$ (defined in \ref{Ia}), we define a covariant
derivative as
\begin{equation}
\nabla_\mu
X^{\nu}=\partial_{\mu}X^{\nu}+\Gamma^{\nu}_{\lambda\mu}X^{\lambda}
\,,\label{3.4a}
\end{equation}
where $\Gamma^{\nu}_{\lambda\mu}$ is a connection that stipulates
the nature of the galilean space-time.

Let us consider the action of the covariant derivative on a vector
field $X^\mu$, introducing the curvature tensor. Hence we write
\begin{equation}
\nabla_{[\mu}
\nabla_{\lambda]}X^{\nu}=\frac{1}{2}R^{\nu}\,_{\gamma\mu\lambda}X^{\gamma}
\,,\label{3.5a}
\end{equation}
where $R^{\nu}\,_{\gamma\mu\lambda}$ is the curvature tensor in
(4+1) dimensions. We assume that when there is no gravitational
field, the curvature tensor vanishes.

The metric tensor is a covariant tensor of rank 2. It is used to
define distances and lengths of vectors. The infinitesimal distance
between two points  $x^{\mu}$ and $x^{\mu}+dx^{\mu}$ in the curved
manifold defined from ${\cal S}_{4,1}$ is
\begin{equation}
ds^2=g_{\mu\nu}dx^{\mu}dx^{\nu}\,,\label{3.6a}
\end{equation}
where $g_{\mu\nu}$ is the metric tensor. The relation given in
Eq.~(\ref{3.6a}) represents the line element as well. We emphasize
that the metric $\eta_{\mu\nu}=diag(1,1,1,-1,1)$ defines a flat line
element. The imposition that the covariant derivative of metric is
zero yields the following expression for the connection
\begin{equation}
\Gamma^{\nu}_{\lambda\mu}=\frac{1}{2}g^{\nu\delta}(\partial_{\lambda}
g_{\delta\mu}+\partial_{\mu}g_{\delta\lambda}-\partial_{\delta}g_{\lambda\mu})\,.\label{3.7a}
\end{equation}
In this case the manifold is said to be affine and the curvature
tensor satisfies the following properties:
\begin{eqnarray}
R_{\mu\nu\lambda\gamma}=-R_{\mu\nu\gamma\lambda}=-R_{\nu\mu\lambda\gamma}&=&
R_{\lambda\gamma\mu\nu}\nonumber \\
R_{\mu\nu\lambda\gamma}+R_{\mu\gamma\nu\lambda}+R_{\mu\lambda\gamma\nu}&\equiv&0.\label{3.8a}
\end{eqnarray}
These properties are derived from Eq.~(\ref{3.5a}). If we perform a
contraction of the indices of the curvature tensor then it is
possible to define the Galilei-invariant curvature scalar
\begin{equation}
R=g^{\mu\nu}g^{\gamma\lambda}R_{\gamma\mu\lambda\nu}.\label{3.9a}
\end{equation}

The field equations are derived from a Lagrangian invariant under
galilean transformations. A natural candidate is the curvature
scalar, that gives rise to the action
\begin{equation}
I=\int_\Omega d\Omega (\sqrt{-g}R+k\mathfrak{L}_m)\,,\label{3.10a}
\end{equation}
where $g=det g_{\mu\nu}$, $k$ is the coupling constant,
$\mathfrak{L}_m$ is a matter lagrangian density and $d\Omega$ is the
5-dimensional element of volume. Varying the action with respect to
$g_{\mu\nu}$, we obtain
\begin{equation}
R^{\mu\nu}-\frac{1}{2}g^{\mu\nu}R=kT^{\mu\nu}\,,\label{3.12a}
\end{equation}
where $T^{\mu\nu}$ is the energy-momentum tensor of matter fields
and $R_{\mu\nu}=R^{\lambda}\,_{\mu\lambda\nu}$. These equations have
the same form as those ones describing the general relativity
equations in (3+1)-Lorentz space-time. Here, however,
Eq.~(\ref{3.12a}) expresses a (3+1)-galilean-transformation
invariant gravity theory when analyzed in the light-cone. The
quantity $k$ is the coupling constant between the galilean gravity
and matter fields. Let us consider an example.

Considering a spherically symmetric line element, we obtain a
geodesic motion by analyzing the functional variation of
\begin{equation}
K=g_{\mu\nu}\dot{x}^{\mu}\dot{x}^{\nu}\,,\label{3.13a}
\end{equation}
which takes the possible values 1, -1 or 0. This relation is similar
to the galilean mass-shell condition, Eq.~(\ref{seg01.111}), and
leads to
\begin{equation}
U^{''}+ U=G\frac{M}{h^2}+ 3\frac{G}{c^{2}}MU^2 \,,\label{4.10a}
\end{equation}
where $U$ is the function that describes the trajectory and the
prime stands for the derivative with respect to the angle
$\phi$~\cite{Ulhoa:2009at}. Then  we have derived, for instance, the
same value for the advance of the Mercury perihelion, as it is
obtained in general relativity considering the post-newtonian limit.
From this result, we conclude that the galilean gravity theory is,
physically, a covariant version of the weak-field limit of the
general relativity. This results is appealing, in particular, for
experimental purposes and represents a prescription to find such a
limit~\cite{Ulhoa:2009at}.

\bibliographystyle{apsrev4-1}
\bibliography{ref}

%Merlin.mbs v4.21 2009-07-09.
\begin{thebibliography}{10}%
\makeatletter
\providecommand \@ifxundefined [1]{%
 \ifx #1\undefined \expandafter \@firstoftwo
 \else \expandafter \@secondoftwo
\fi
}%
\providecommand \@ifnum [1]{%
 \ifnum #1\expandafter \@firstoftwo
 \else \expandafter \@secondoftwo
\fi
}%
\providecommand \enquote [1]{``#1''}%
\providecommand \bibnamefont  [1]{#1}%
\providecommand \bibfnamefont [1]{#1}%
\providecommand \citenamefont [1]{#1}%
\providecommand\href[0]{\@sanitize\@href}%
\providecommand\@href[1]{\endgroup\@@startlink{#1}\endgroup\@@href}%
\providecommand\@@href[1]{#1\@@endlink}%
\providecommand \@sanitize [0]{\begingroup\catcode`\&12\catcode`\#12\relax}%
\@ifxundefined \pdfoutput {\@firstoftwo}{%
 \@ifnum{\z@=\pdfoutput}{\@firstoftwo}{\@secondoftwo}%
}{%
 \providecommand\@@startlink[1]{\leavevmode\special{html:<a href="#1">}}%
 \providecommand\@@endlink[0]{\special{html:</a>}}%
}{%
 \providecommand\@@startlink[1]{%
  \leavevmode
  \pdfstartlink
   attr{/Border[0 0 1 ]/H/I/C[0 1 1]}%
   user{/Subtype/Link/A<</Type/Action/S/URI/URI(#1)>>}%
  \relax
 }%
 \providecommand\@@endlink[0]{\pdfendlink}%
}%
\providecommand \url  [0]{\begingroup\@sanitize \@url }%
\providecommand \@url [1]{\endgroup\@href {#1}{\urlprefix}}%
\providecommand \urlprefix [0]{URL }%
\providecommand \Eprint[0]{\href }%
\@ifxundefined \urlstyle {%
  \providecommand \doi [1]{doi:\discretionary{}{}{}#1}%
}{%
  \providecommand \doi [0]{doi:\discretionary{}{}{}\begingroup
  \urlstyle{rm}\Url }%
}%
\providecommand \doibase [0]{http://dx.doi.org/}%
\providecommand \Doi[1]{\href{\doibase#1}}%
\providecommand \bibAnnote [3]{%
  \BibitemShut{#1}%
  \begin{quotation}\noindent
    \textsc{Key:}\ #2\\\textsc{Annotation:}\ #3%
  \end{quotation}%
}%
\providecommand \bibAnnoteFile [2]{%
  \IfFileExists{#2}{\bibAnnote {#1} {#2} {\input{#2}}}{}%
}%
\providecommand \typeout [0]{\immediate \write \m@ne }%
\providecommand \selectlanguage [0]{\@gobble}%
\providecommand \bibinfo [0]{\@secondoftwo}%
\providecommand \bibfield [0]{\@secondoftwo}%
\providecommand \translation [1]{[#1]}%
\providecommand \BibitemOpen[0]{}%
\providecommand \bibitemStop [0]{}%
\providecommand \bibitemNoStop [0]{.\EOS\space}%
\providecommand \EOS [0]{\spacefactor3000\relax}%
\providecommand \BibitemShut [1]{\csname bibitem#1\endcsname}%
%</preamble>
\bibitem{Fetter}%
  \BibitemOpen
  \bibfield{author}{%
  \bibinfo {author} {\bibfnamefont{A.~L.}\ \bibnamefont{Fetter}}\ and\ \bibinfo
  {author} {\bibfnamefont{J.~D.}\ \bibnamefont{Walecka}},\ }%
  \emph{\bibinfo {title} {Quantum Theory of Many-Particle Systems}}\ (\bibinfo
  {publisher} {Dover Publications},\ \bibinfo {year} {2003})\ ISBN \bibinfo
  {isbn} {0486428273}%
  \bibAnnoteFile{NoStop}{Fetter}%
\bibitem{PhysRev.150.1313}%
  \BibitemOpen
  \bibfield{author}{%
  \bibinfo {author} {\bibfnamefont{S.}~\bibnamefont{Weinberg}},\ }%
  \bibfield{journal}{%
  \Doi{10.1103/PhysRev.150.1313}{\bibinfo {journal} {Phys. Rev.}}\ }%
  \textbf{\bibinfo {volume} {150}},\ \bibinfo {pages} {1313} (\bibinfo {month}
  {Oct}\ \bibinfo {year} {1966})%
  \bibAnnoteFile{NoStop}{PhysRev.150.1313}%
\bibitem{PhysRev.165.1535}%
  \BibitemOpen
  \bibfield{author}{%
  \bibinfo {author} {\bibfnamefont{L.}~\bibnamefont{Susskind}},\ }%
  \bibfield{journal}{%
  \Doi{10.1103/PhysRev.165.1535}{\bibinfo {journal} {Phys. Rev.}}\ }%
  \textbf{\bibinfo {volume} {165}},\ \bibinfo {pages} {1535} (\bibinfo {month}
  {Jan}\ \bibinfo {year} {1968})%
  \bibAnnoteFile{NoStop}{PhysRev.165.1535}%
\bibitem{PhysRevD.1.2795}%
  \BibitemOpen
  \bibfield{author}{%
  \bibinfo {author} {\bibfnamefont{D.}~\bibnamefont{Flory}},\ }%
  \bibfield{journal}{%
  \Doi{10.1103/PhysRevD.1.2795}{\bibinfo {journal} {Phys. Rev. D}}\ }%
  \textbf{\bibinfo {volume} {1}},\ \bibinfo {pages} {2795} (\bibinfo {month}
  {May}\ \bibinfo {year} {1970})%
  \bibAnnoteFile{NoStop}{PhysRevD.1.2795}%
\bibitem{PhysRevD.1.2901}%
  \BibitemOpen
  \bibfield{author}{%
  \bibinfo {author} {\bibfnamefont{J.~B.}\ \bibnamefont{Kogut}}\ and\ \bibinfo
  {author} {\bibfnamefont{D.~E.}\ \bibnamefont{Soper}},\ }%
  \bibfield{journal}{%
  \Doi{10.1103/PhysRevD.1.2901}{\bibinfo {journal} {Phys. Rev. D}}\ }%
  \textbf{\bibinfo {volume} {1}},\ \bibinfo {pages} {2901} (\bibinfo {month}
  {May}\ \bibinfo {year} {1970})%
  \bibAnnoteFile{NoStop}{PhysRevD.1.2901}%
\bibitem{Beckers1975338}%
  \BibitemOpen
  \bibfield{author}{%
  \bibinfo {author} {\bibfnamefont{J.}~\bibnamefont{Beckers}}\ and\ \bibinfo
  {author} {\bibfnamefont{M.}~\bibnamefont{Jaspers}},\ }%
  \bibfield{journal}{%
  \Doi{DOI: 10.1016/0378-4371(75)90032-1}{\bibinfo {journal} {Physica A:
  Statistical and Theoretical Physics}}\ }%
  \textbf{\bibinfo {volume} {79}},\ \bibinfo {pages} {338 } (\bibinfo {year}
  {1975}),\ ISSN \bibinfo {issn} {0378-4371},\
  \url{http://www.sciencedirect.com/science/article/B6TVG-46CC9SB-2J/2/336f249%
e1e3b98c668e0e78f91a8993f}%
  \bibAnnoteFile{NoStop}{Beckers1975338}%
\bibitem{PhysRevD.52.5954}%
  \BibitemOpen
  \bibfield{author}{%
  \bibinfo {author} {\bibfnamefont{N.~E.}\ \bibnamefont{Ligterink}}\ and\
  \bibinfo {author} {\bibfnamefont{B.~L.~G.}\ \bibnamefont{Bakker}},\ }%
  \bibfield{journal}{%
  \Doi{10.1103/PhysRevD.52.5954}{\bibinfo {journal} {Phys. Rev. D}}\ }%
  \textbf{\bibinfo {volume} {52}},\ \bibinfo {pages} {5954} (\bibinfo {month}
  {Nov}\ \bibinfo {year} {1995})%
  \bibAnnoteFile{NoStop}{PhysRevD.52.5954}%
\bibitem{PhysRevD.57.4965}%
  \BibitemOpen
  \bibfield{author}{%
  \bibinfo {author} {\bibfnamefont{N.~C.~J.}\ \bibnamefont{Schoonderwoerd}}\
  and\ \bibinfo {author} {\bibfnamefont{B.~L.~G.}\ \bibnamefont{Bakker}},\ }%
  \bibfield{journal}{%
  \Doi{10.1103/PhysRevD.57.4965}{\bibinfo {journal} {Phys. Rev. D}}\ }%
  \textbf{\bibinfo {volume} {57}},\ \bibinfo {pages} {4965} (\bibinfo {month}
  {Apr}\ \bibinfo {year} {1998})%
  \bibAnnoteFile{NoStop}{PhysRevD.57.4965}%
\bibitem{PhysRevD.58.025013}%
  \BibitemOpen
  \bibfield{author}{%
  \bibinfo {author} {\bibfnamefont{N.~C.~J.}\ \bibnamefont{Schoonderwoerd}}\
  and\ \bibinfo {author} {\bibfnamefont{B.~L.~G.}\ \bibnamefont{Bakker}},\ }%
  \bibfield{journal}{%
  \Doi{10.1103/PhysRevD.58.025013}{\bibinfo {journal} {Phys. Rev. D}}\ }%
  \textbf{\bibinfo {volume} {58}},\ \bibinfo {pages} {025013} (\bibinfo {month}
  {Jun}\ \bibinfo {year} {1998})%
  \bibAnnoteFile{NoStop}{PhysRevD.58.025013}%
\bibitem{PhysRevD.62.105015}%
  \BibitemOpen
  \bibfield{author}{%
  \bibinfo {author} {\bibfnamefont{A.}~\bibnamefont{Harindranath}}, \bibinfo
  {author} {\bibfnamefont{L.}~\bibnamefont{Martinovic}},\ and\ \bibinfo
  {author} {\bibfnamefont{J.~P.}\ \bibnamefont{Vary}},\ }%
  \bibfield{journal}{%
  \Doi{10.1103/PhysRevD.62.105015}{\bibinfo {journal} {Phys. Rev. D}}\ }%
  \textbf{\bibinfo {volume} {62}},\ \bibinfo {pages} {105015} (\bibinfo {month}
  {Oct}\ \bibinfo {year} {2000})%
  \bibAnnoteFile{NoStop}{PhysRevD.62.105015}%
\bibitem{Montigny}%
  \BibitemOpen
  \bibfield{author}{%
  \bibinfo {author} {\bibfnamefont{M.}~\bibnamefont{de~Montigny}}, \bibinfo
  {author} {\bibfnamefont{F.~C.}\ \bibnamefont{Khanna}}, \bibinfo {author}
  {\bibfnamefont{A.~E.}\ \bibnamefont{Santana}},\ and\ \bibinfo {author}
  {\bibfnamefont{E.~S.}\ \bibnamefont{Santos}},\ }%
  \bibfield{journal}{%
  \bibinfo {journal} {J. Phys. A: Math. Gen.}\ }%
  \textbf{\bibinfo {volume} {34}},\ \bibinfo {pages} {8901} (\bibinfo {year}
  {2001})%
  \bibAnnoteFile{NoStop}{Montigny}%
\bibitem{Brihaye}%
  \BibitemOpen
  \bibfield{author}{%
  \bibinfo {author} {\bibfnamefont{Y.}~\bibnamefont{Brihaye}}, \bibinfo
  {author} {\bibfnamefont{C.}~\bibnamefont{Gonera}}, \bibinfo {author}
  {\bibfnamefont{S.}~\bibnamefont{Giller}},\ and\ \bibinfo {author}
  {\bibfnamefont{P.}~\bibnamefont{Kosinski}}}%
   (\bibinfo {year} {1995}),\
  \Eprint{http://arxiv.org/abs/hep-th/9503046}{arXiv:hep-th/9503046}%
  \bibAnnoteFile{NoStop}{Brihaye}%
%%CITATION = HEP-TH/9503046;%%
\bibitem{Hagen}%
  \BibitemOpen
  \bibfield{author}{%
  \bibinfo {author} {\bibfnamefont{C.~R.}\ \bibnamefont{Hagen}},\ }%
  \bibfield{journal}{%
  \Doi{10.1103/PhysRevD.5.377}{\bibinfo {journal} {Phys. Rev. D}}\ }%
  \textbf{\bibinfo {volume} {5}},\ \bibinfo {pages} {377} (\bibinfo {month}
  {Jan}\ \bibinfo {year} {1972})%
  \bibAnnoteFile{NoStop}{Hagen}%
\bibitem{conf1}%
  \BibitemOpen
  \bibfield{author}{%
  \bibinfo {author} {\bibfnamefont{Y.}~\bibnamefont{Nishida}}\ and\ \bibinfo
  {author} {\bibfnamefont{D.~T.}\ \bibnamefont{Son}},\ }%
  \bibfield{journal}{%
  \Doi{10.1103/PhysRevD.76.086004}{\bibinfo {journal} {Phys. Rev. D}}\ }%
  \textbf{\bibinfo {volume} {76}},\ \bibinfo {pages} {086004} (\bibinfo {month}
  {Oct}\ \bibinfo {year} {2007})%
  \bibAnnoteFile{NoStop}{conf1}%
\bibitem{conf2}%
  \BibitemOpen
  \bibfield{author}{%
  \bibinfo {author} {\bibfnamefont{K.}~\bibnamefont{Balasubramanian}}\ and\
  \bibinfo {author} {\bibfnamefont{J.}~\bibnamefont{McGreevy}},\ }%
  \bibfield{journal}{%
  \Doi{10.1103/PhysRevLett.101.061601}{\bibinfo {journal} {Phys. Rev. Lett.}}\
  }%
  \textbf{\bibinfo {volume} {101}},\ \bibinfo {pages} {061601} (\bibinfo
  {month} {Aug}\ \bibinfo {year} {2008})%
  \bibAnnoteFile{NoStop}{conf2}%
\bibitem{conf3}%
  \BibitemOpen
  \bibfield{author}{%
  \bibinfo {author} {\bibfnamefont{D.~T.}\ \bibnamefont{Son}},\ }%
  \bibfield{journal}{%
  \Doi{10.1103/PhysRevD.78.046003}{\bibinfo {journal} {Phys. Rev. D}}\ }%
  \textbf{\bibinfo {volume} {78}},\ \bibinfo {pages} {046003} (\bibinfo {month}
  {Aug}\ \bibinfo {year} {2008})%
  \bibAnnoteFile{NoStop}{conf3}%
\bibitem{Khanna}%
  \BibitemOpen
  \bibfield{author}{%
  \bibinfo {author} {\bibfnamefont{F.~C.}\ \bibnamefont{Khanna}}, \bibinfo
  {author} {\bibfnamefont{A.~E.}\ \bibnamefont{Santana}}, \bibinfo {author}
  {\bibfnamefont{A.}~\bibnamefont{Matos~Neto}}, \bibinfo {author}
  {\bibfnamefont{J.~D.~M.}\ \bibnamefont{Vianna}},\ and\ \bibinfo {author}
  {\bibfnamefont{T.}~\bibnamefont{Kopf}}}%
   (\bibinfo {year} {1998}),\
  \Eprint{http://arxiv.org/abs/hep-th/9812222}{arXiv:hep-th/9812222}%
  \bibAnnoteFile{NoStop}{Khanna}%
%%CITATION = HEP-TH/9812222;%%
\bibitem{Santos}%
  \BibitemOpen
  \bibfield{author}{%
  \bibinfo {author} {\bibfnamefont{E.~S.}\ \bibnamefont{Santos}}, \bibinfo
  {author} {\bibfnamefont{M.}~\bibnamefont{de~Montigny}}, \bibinfo {author}
  {\bibfnamefont{F.~C.}\ \bibnamefont{Khanna}},\ and\ \bibinfo {author}
  {\bibfnamefont{A.~E.}\ \bibnamefont{Santana}},\ }%
  \bibfield{journal}{%
  \bibinfo {journal} {J. Phys. A: Math. Gen.}\ }%
  \textbf{\bibinfo {volume} {37}},\ \bibinfo {pages} {9771} (\bibinfo {year}
  {2004})%
  \bibAnnoteFile{NoStop}{Santos}%
\bibitem{Ulhoa:2009at}%
  \BibitemOpen
  \bibfield{author}{%
  \bibinfo {author} {\bibfnamefont{S.~C.}\ \bibnamefont{Ulhoa}}, \bibinfo
  {author} {\bibfnamefont{F.~C.}\ \bibnamefont{Khanna}},\ and\ \bibinfo
  {author} {\bibfnamefont{A.~E.}\ \bibnamefont{Santana}},\ }%
  \bibfield{journal}{%
  \Doi{10.1142/S0217751X09046333}{\bibinfo {journal} {Int. J. Mod. Phys.}}\ }%
  \textbf{\bibinfo {volume} {A24}},\ \bibinfo {pages} {5287} (\bibinfo {year}
  {2009}),\ \Eprint{http://arxiv.org/abs/0902.2023}{arXiv:0902.2023 [gr-qc]}%
  \bibAnnoteFile{NoStop}{Ulhoa:2009at}%
%%CITATION = 0902.2023;%%
\bibitem{Cuzinatto:2009dw}%
  \BibitemOpen
  \bibfield{author}{%
  \bibinfo {author} {\bibfnamefont{R.}~\bibnamefont{Cuzinatto}}, \bibinfo
  {author} {\bibfnamefont{P.}~\bibnamefont{Pompeia}}, \bibinfo {author}
  {\bibfnamefont{M.}~\bibnamefont{de~Montigny}},\ and\ \bibinfo {author}
  {\bibfnamefont{F.}~\bibnamefont{Khanna}},\ }%
  \bibfield{journal}{%
  \Doi{DOI: 10.1016/j.physletb.2009.08.024}{\bibinfo {journal} {Physics Letters
  B}}\ }%
  \textbf{\bibinfo {volume} {680}},\ \bibinfo {pages} {98 } (\bibinfo {year}
  {2009}),\ ISSN \bibinfo {issn} {0370-2693},\
  \url{http://www.sciencedirect.com/science/article/B6TVN-4X0F3RJ-6/2/e9964fba%
a80bc2e98f816ab1e576533c}%
  \bibAnnoteFile{NoStop}{Cuzinatto:2009dw}%
\bibitem{Hehl2}%
  \BibitemOpen
  \bibfield{author}{%
  \bibinfo {author} {\bibfnamefont{F.~W.}\ \bibnamefont{Hehl}}, \bibinfo
  {author} {\bibfnamefont{J.}~\bibnamefont{Lemke}},\ and\ \bibinfo {author}
  {\bibfnamefont{E.~W.}\ \bibnamefont{Mielke}},\ }%
  in\ \emph{\bibinfo {booktitle} {Geometry and Theoretical Physics}},\ \bibinfo
  {editor} {edited by\ \bibinfo {editor}
  {\bibfnamefont{J.}~\bibnamefont{Debrus}}\ and\ \bibinfo {editor}
  {\bibfnamefont{A.~C.}\ \bibnamefont{Hirshfeld}}}\ (\bibinfo {publisher}
  {Springer},\ \bibinfo {address} {Berlin},\ \bibinfo {year} {1991})%
  \bibAnnoteFile{NoStop}{Hehl2}%
\bibitem{hehl-1994}%
  \BibitemOpen
  \bibfield{author}{%
  \bibinfo {author} {\bibfnamefont{F.~W.}\ \bibnamefont{Hehl}}, \bibinfo
  {author} {\bibfnamefont{J.~D.}\ \bibnamefont{McCrea}}, \bibinfo {author}
  {\bibfnamefont{E.~W.}\ \bibnamefont{Mielke}},\ and\ \bibinfo {author}
  {\bibfnamefont{Y.}~\bibnamefont{Ne'eman}},\ }%
  \enquote{\bibinfo {title} {Metric-affine gauge theory of gravity: Field
  equations, noether identities, world spinors, and breaking of dilation
  invariance},}\  (\bibinfo {year} {1994})%
  \bibAnnoteFile{NoStop}{hehl-1994}%
\bibitem{Maluf2}%
  \BibitemOpen
  \bibfield{author}{%
  \bibinfo {author} {\bibfnamefont{J.~W.}\ \bibnamefont{Maluf}},\ }%
  \bibfield{journal}{%
  \Doi{10.1002/andp.200510161}{\bibinfo {journal} {Annalen Phys.}}\ }%
  \textbf{\bibinfo {volume} {14}},\ \bibinfo {pages} {723} (\bibinfo {year}
  {2005}),\ \Eprint{http://arxiv.org/abs/gr-qc/0504077}{arXiv:gr-qc/0504077}%
  \bibAnnoteFile{NoStop}{Maluf2}%
%%CITATION = GR-QC/0504077;%%
\bibitem{MalufS}%
  \BibitemOpen
  \bibfield{author}{%
  \bibinfo {author} {\bibfnamefont{J.}~\bibnamefont{Maluf}}, \bibinfo {author}
  {\bibfnamefont{S.}~\bibnamefont{Ulhoa}}, \bibinfo {author}
  {\bibfnamefont{F.}~\bibnamefont{Faria}},\ and\ \bibinfo {author}
  {\bibfnamefont{J.}~\bibnamefont{da~Rocha-Neto}},\ }%
  \bibfield{journal}{%
  \bibinfo {journal} {Classical and Quantum Gravity}\ }%
  \textbf{\bibinfo {volume} {23}},\ \bibinfo {pages} {6245} (\bibinfo {year}
  {2006}),\ \url{http://stacks.iop.org/0264-9381/23/6245}%
  \bibAnnoteFile{NoStop}{MalufS}%
\bibitem{Aldrovandi}%
  \BibitemOpen
  \bibfield{author}{%
  \bibinfo {author} {\bibfnamefont{R.}~\bibnamefont{Aldrovandi}}, \bibinfo
  {author} {\bibfnamefont{J.~G.}\ \bibnamefont{Pereira}},\ and\ \bibinfo
  {author} {\bibfnamefont{K.~H.}\ \bibnamefont{Vu}},\ }%
  \bibfield{journal}{%
  \bibinfo {journal} {Braz. J. Phys.}\ }%
  \textbf{\bibinfo {volume} {34}},\ \bibinfo {pages} {1374} (\bibinfo {year}
  {2004}),\ \Eprint{http://arxiv.org/abs/gr-qc/0312008}{arXiv:gr-qc/0312008}%
  \bibAnnoteFile{NoStop}{Aldrovandi}%
%%CITATION = GR-QC/0312008;%%
\bibitem{Babak}%
  \BibitemOpen
  \bibfield{author}{%
  \bibinfo {author} {\bibfnamefont{S.~V.}\ \bibnamefont{Babak}}\ and\ \bibinfo
  {author} {\bibfnamefont{L.~P.}\ \bibnamefont{Grishchuk}},\ }%
  \bibfield{journal}{%
  \Doi{10.1103/PhysRevD.61.024038}{\bibinfo {journal} {Phys. Rev. D}}\ }%
  \textbf{\bibinfo {volume} {61}},\ \bibinfo {pages} {024038} (\bibinfo {month}
  {Dec}\ \bibinfo {year} {1999})%
  \bibAnnoteFile{NoStop}{Babak}%
\bibitem{Dinverno}%
  \BibitemOpen
  \bibfield{author}{%
  \bibinfo {author} {\bibfnamefont{R.}~\bibnamefont{d'Inverno}},\ }%
  \emph{\bibinfo {title} {Introducing Einstein's Relativity}},\ \bibinfo
  {edition} {4th}\ ed.\ (\bibinfo {publisher} {Clarendon Press, Oxford},\
  \bibinfo {year} {1996})%
  \bibAnnoteFile{NoStop}{Dinverno}%
\bibitem{Landau}%
  \BibitemOpen
  \bibfield{author}{%
  \bibinfo {author} {\bibfnamefont{L.~D.}\ \bibnamefont{Landau}}\ and\ \bibinfo
  {author} {\bibfnamefont{E.~M.}\ \bibnamefont{Lifshiz}},\ }%
  \emph{\bibinfo {title} {The Classical Theory of Fields}},\ \bibinfo {edition}
  {4th}\ ed.,\ \bibinfo {series} {Course of Theoretical Physics}, Vol.~\bibinfo
  {volume} {2}\ (\bibinfo {publisher} {Elsevier Butterworth-Heinemann},\
  \bibinfo {year} {2004})%
  \bibAnnoteFile{NoStop}{Landau}%
\bibitem{Kobayashi:2007jn}%
  \BibitemOpen
  \bibfield{author}{%
  \bibinfo {author} {\bibfnamefont{M.}~\bibnamefont{Kobayashi}}, \bibinfo
  {author} {\bibfnamefont{M.}~\bibnamefont{de~Montigny}},\ and\ \bibinfo
  {author} {\bibfnamefont{F.~C.}\ \bibnamefont{Khanna}},\ }%
  \bibfield{journal}{%
  \Doi{10.1088/1751-8113/41/12/125402}{\bibinfo {journal} {J. Phys.}}\ }%
  \textbf{\bibinfo {volume} {A41}},\ \bibinfo {pages} {125402} (\bibinfo {year}
  {2008}),\ \Eprint{http://arxiv.org/abs/0710.5556}{arXiv:0710.5556 [hep-th]}%
  \bibAnnoteFile{NoStop}{Kobayashi:2007jn}%
%%CITATION = 0710.5556;%%
\bibitem{Montigny:2007bv}%
  \BibitemOpen
  \bibfield{author}{%
  \bibinfo {author} {\bibfnamefont{M.}~\bibnamefont{de~Montigny}}, \bibinfo
  {author} {\bibfnamefont{F.~C.}\ \bibnamefont{Khanna}},\ and\ \bibinfo
  {author} {\bibfnamefont{F.~M.}\ \bibnamefont{Saradzhev}},\ }%
  \bibfield{journal}{%
  \Doi{10.1016/j.aop.2007.08.002}{\bibinfo {journal} {Annals Phys.}}\ }%
  \textbf{\bibinfo {volume} {323}},\ \bibinfo {pages} {1191} (\bibinfo {year}
  {2008}),\ \Eprint{http://arxiv.org/abs/0706.4106}{arXiv:0706.4106 [hep-th]}%
  \bibAnnoteFile{NoStop}{Montigny:2007bv}%
%%CITATION = 0706.4106;%%
\bibitem{1978GReGr...9..621C}%
  \BibitemOpen
  \bibfield{author}{%
  \bibinfo {author} {\bibfnamefont{N.~C.~T.}\ \bibnamefont{{Coote}}}\ and\
  \bibinfo {author} {\bibfnamefont{A.~J.}\ \bibnamefont{{Macfarlane}}},\ }%
  \bibfield{journal}{%
  \Doi{10.1007/BF00761006}{\bibinfo {journal} {General Relativity and
  Gravitation}}\ }%
  \textbf{\bibinfo {volume} {9}},\ \bibinfo {pages} {621} (\bibinfo {month}
  {Jul.}\ \bibinfo {year} {1978})%
  \bibAnnoteFile{NoStop}{1978GReGr...9..621C}%
\end{thebibliography}%

\end{document}